\documentclass[ aps,
                pra,
                showkeys,
                reprint,
                superscriptaddress,
                floatfix
                ]{revtex4-1}

\usepackage[english]{babel}
\usepackage{graphicx}
\usepackage{dcolumn}
\usepackage{bm}
\usepackage{braket}
\usepackage{amsmath,amssymb}
\usepackage{hyperref} 

\newcommand{\ie}{i.\,e, }

\newcommand{\Eg}{E.\,g, } 

\begin{document}
\preprint{Bell/from/Hell/2.0}

\title[Local Classical Strategies vs Geometrical Quantum Constraints]{Local Classical Strategies vs Geometrical Quantum Constraints}

\author{R. Restrepo-Villegas}
\email{r.restrepovillegas@gmail.com}
\affiliation{Grupo de F\'isica At\'omica y Molecular (GFAM), Instituto de F\'isica, Facultad de Ciencias Exactas y Naturales, Universidad de Antioquia UdeA, Calle 70 No. 52-21, Medell\'in, Colombia.} 

\author{E. Agudelo}
\affiliation{Arbeitsgruppe Theoretische Quantenoptik, Institut f\"ur Physik, Universit\"at Rostock, D-18051 Rostock, Germany.}

\author{J. Castrill\'on}
\affiliation{Grupo de F\'isica At\'omica y Molecular (GFAM), Instituto de F\'isica, Facultad de Ciencias Exactas y Naturales, Universidad de Antioquia UdeA, Calle 70 No. 52-21, Medell\'in, Colombia.} 

\author{B. A. Rodr\'iguez}%
\email{boris.rodriguez@udea.edu.co}
\affiliation{Grupo de F\'isica At\'omica y Molecular (GFAM), Instituto de F\'isica, Facultad de Ciencias Exactas y Naturales, Universidad de Antioquia UdeA, Calle 70 No. 52-21, Medell\'in, Colombia.} 
\date{\today}

\begin{abstract}
We use an alternative approach to show that quantum entanglement-like correlations cannot be reproduced for any classical protocol. 
In our proposal, quantum geometric restrictions are impose over the physical system related to the existence of entanglement and we demonstrate that there is no classical local strategy that can reproduce them completely. 
Typically, the implementations of Bell inequalities have as a starting point the expectation of classical behavior and as conclusion the violation due to the quantum character of the system.
We go the other way around. For this purpose, we build a computational simulation based on the scheme of non-communicating students. 
In this scheme, the students cannot manipulate the quantum systems but they may set up in advance a common strategy and share some common classical data in order to try to reproduce the given quantum correlations of such systems. 
By thoroughly searching in the whole space of classical strategies we conclude that local operations and classical communications does not satisfy the geometrical constraints imposed by quantum entanglement. 
\end{abstract}

\maketitle

\section{\label{sec1}Introduction}
Quantum entanglement is, without a doubt, one of the most interesting phenomena of quantum physics. 
It describes the event in which two (or more) systems can be intimately connected to each other, no matter how far they are. 
Moreover, there is no classical equivalent to quantum entanglement, that is, it is a purely quantum phenomenon. 
Entanglement was a term originally conceived by E. Schr\"odinger~\cite{Schrodinger1935} to name non-local correlations between physical systems with quantum properties and used audaciously by A. Einstein, B. Podolsky and N. Rosen (EPR) in their 1935 famous paper, in an attempt to demonstrate that Quantum Mechanics (QM) was an incomplete theory~\cite{epr}. 
For EPR it was inadmissible that the properties of remote systems can be intrinsically linked as predicted by the QM.
Supposing that the properties of  the composing parts of a physical system define the system a priori in its totality is a premise that can be classified as self-evident, at least classically speaking. 
During the early years of the first quantum revolution, this kind of discussions were relegated  to old men and philosophers because apparently such debate had nothing to do with physics and it was merely a matter of interpretation. 
It was J. S. Bell who in 1964 proposed a set of inequalities that could be realized experimentally with present technology~\cite{Bell1964}. 
Bell's ideas changed our classic and naive notion of reality forever. 
As noted by N. D. Mermin: ``The point is no longer that quantum mechanics is an extraordinarily peculiar theory, but that the world is an extraordinarily peculiar place.''~\cite{Mermin1981}. 
Soon afterwards, experiments~\cite{Aspect1982, Zeilinger1998, Zeilinger2015, Hensen2015, Aspect2015, Zeilinger2017} showed that entanglement and therefore QM, fully describe objective physical reality. 

There are different ways to propose the Bell's theorem, or his inequalities~\cite{CHSH1969, Clauser1978, Mermin1990, GHZ1990, Mermin1994, Banghelo2016}. 
Generally, one assumes the suppositions of local realism and show that any prediction based on these assumptions disagrees with the predictions made by the QM, thus representing a violation of Bell's inequalities and local realism by quantum theory. 
We present here an alternative approach to show that quantum entanglement-like correlations cannot be reproduced for any classical mean.
We start from quantum predictions, more specifically, from geometric constraints imposed by entanglement, and we show that there is no possible classical scenario capable of satisfying such predictions completely. 
It is our aim to argue that protocols involving only local operations and classical communications (LOCC) cannot replicate the behavior of a quantum entangled system. 
We discuss our findings in particular in the case of pairs of entangled photons in polarization.
The method we propose, consists on a exhaustive scan over a space of parameters that characterizes the different classic strategies that can be considered in order to simulate the probabilities of coincidences obtained in a quantum experiment with photon pairs prepared in a maximally entangled state.
We show how none of the conceivable strategies can produce solutions out of the region defined by the classical conditions.

The paper is organized as follows. 
We start with the description of the problem and the experimental protocols to be compared in Sec.~{\ref{sec2}}, the quantum and the classical one. 
In Sec.~\ref{sec3} we discuss the different initial conditions treated together with its probabilities of coincidence. 
We compare the predictions of quantum theory with the classical results obtained by the non-communicating students protocol, in the form of a numerical simulation that investigates all the possible strategies them could agree on. 
Finally, in Sec.~\ref{sec4} we discuss our findings and present the conclusions of this work.

\section{\label{sec2} Quantum experiment vs classical simulation}
In order to address the question whether a probability distribution obtained through the performance of an experiment with quantically correlated particles can be classically simulated, we consider the following two scenarios. 
Firstly, we deal with a quantum experiment based on polarization measurements of entangled photon pairs, and secondly, as the classical situation, we consider two uncommunicated students with unlimited classical power and a source of random numbers who will try to simulate the given experimental outcome.

\subsection{\label{sec2:subsec1} Quantum setup: Bell's experiment with photons}
Our quantum experiment is based on the polarization measurements of entangled photon pairs.
Which is a standard technique in quantum optics \cite{clauser1974, Clauser1976,Shih2001a,Shih2001b,Shih2001c,Dehlinger2002}.
Typically, in this scenario a non-linear crystal is pumped with a laser. 
Through Spontaneous Parametric Down-Conversion (SPDC) a single input photon is split into two output photons, called the signal and the idler.
After the pair of photons emerge simultaneously from the common source, the crystal, they are spatially separated and can be put to travel in opposite ways (space-like separated).
Consequently, two distant observers can make measurements of these individual systems.
The two independent observers, Alice and Bob, look for correlations between the polarization of each photon in a pair emitted simultaneously. Such scheme, it is the so-called Bell's experiment with pairs of entangled photons~\cite{Shih2011introduction}.

\begin{figure}
\includegraphics[width=0.95\linewidth]{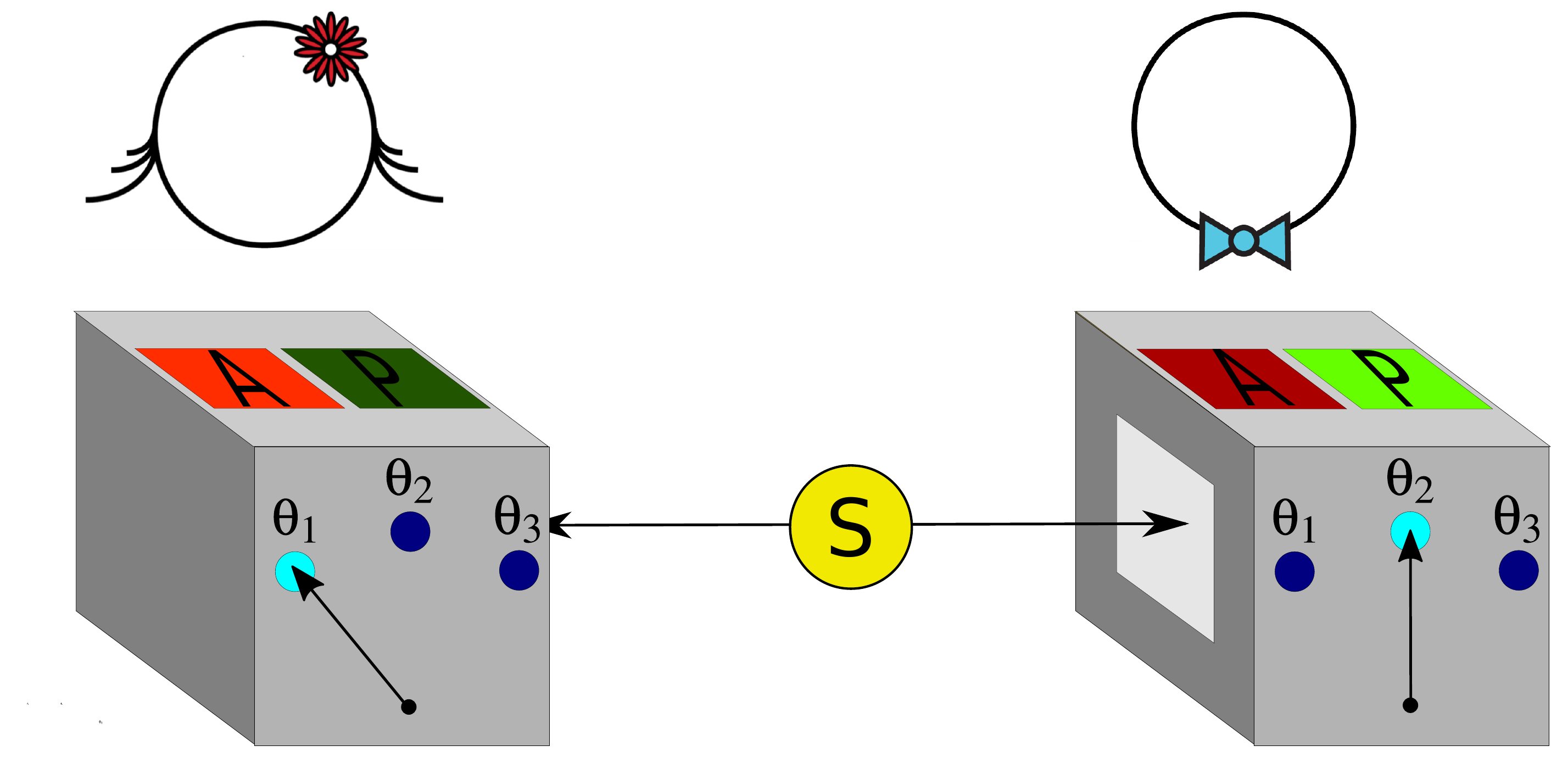}
\caption{Sketch of a Bell experiment made with photons. 
From the source $(S)$ two entangled photons travel in opposite directions until reach two distance observers, Alice and Bob, which have a measuring apparatus each one, such device have a polarization analyzer (PA) and a knob to fix the angle of PA, restricted to three options: $0^{\circ}$, $30^{\circ}$, and $60^{\circ}$. The device have two red $(A)$ and green $(P)$ lights that show if photons pass through of the PA or not. On each run of the experiment, each observer performs a measurement on his own photon. 
The measurement chosen by Alice is labeled $\theta_A$ and its outcome $R_A$. Similarly for Bob, $\theta_B$ is his measurement and $R_B$ its outcome.}
\label{fig1}
\end{figure}

\textit{Polarization measurements---} 
Let us consider the Bell state $\ket{\phi^+} = \frac{1}{\sqrt{2}}\left(\ket{HH} + \ket{VV}\right)$ (maximally entangled) and a type I SPDC. 
Here, the polarization of the two photons is the same, and orthogonal to the polarization of the pump photon. 
It is our aim to describe the correlations between the pair of photons by the measurements of their individual polarization. 
For this purpose, both Alice and Bob have a polarization analyzer (PA), and in each run of the experiment they can select in their own PA in some angles $\theta_A$ and $\theta_B$, respectively, and finally observe the outcomes of each PA.  
There are just two kind of possible outcomes: the photon pass $(P)$ through the PA, or well the photon is absorbed $(A)$ by the PA, cf. Fig.~\ref{fig1}. 
Therefore, for many runs of the experiment, the outcomes $(A)$ and $(P)$ may vary, even if the same measurements are made. 
We labeled $R_A$ and $R_B$ the outcomes of Alice and Bob measurements, respectively. 
In this scenario, a possible experiment could be to set: $\theta_A = 0^{\circ}$ and $\theta_B = 30^{\circ}$; with resulting outcomes $(A)$ and $(P)$, respectively.
Without loss of generality and with the aim of simplifying the experiment and the calculations, we will restrict the measurements to only three angles. 
Such angles are $0^{\circ}$, $30^{\circ}$, and $60^{\circ}$, \ie on each run of the experiment, Alice and Bob will randomly pick one of these three values, set their own PA accordingly, and record the outcome together with the chosen angle.

In case that the photons have been prepared in the particular state $\ket{\phi^+}$, QM predicts that the joint probability of the measurement coincidences on both sides is determined by the square cosine of the phase shift between the angles of both PAs, $(\Delta \theta = |\theta_A - \theta_B|)$.
Now, for the three angles that we chose, the phase shift could only be $\Delta \theta = \{0^{\circ}, 30^{\circ}, 60^{\circ}\}$. Therefore, we have nine possible configurations for the PA pair: $0^{\circ}-0^{\circ}$, $0^{\circ}-30^{\circ}$, $0^{\circ}-60^{\circ}$, $30^{\circ}-0^{\circ}$, $30^{\circ}-30^{\circ}$, $30^{\circ}-60^{\circ}$, $60^{\circ}-0^{\circ}$, $60^{\circ}-30^{\circ}$, and $60^{\circ}-60^{\circ}$.

\textit{Quantum facts---}
Based on the three possible mismatches we can state three characteristics or quantum facts that Alice and Bob observe about the behavior of their photons. 
Under the initial condition, $\ket{\phi^+}$, these three facts are:
\begin{itemize}
\item[$F_1$:]  When the two PAs are oriented at the same angle ($\Delta\theta =0^{\circ}$), the joint probability of coincidence is always equal to unity.
\item[$F_2$:] When the offset at the angle of the PAs is equal to $30^{\circ}$, the joint probability of coincidence is equal to $3/4$.
\item[$F_3$:] When the offset at the angle of the PAs is equal to $60^{\circ}$, the joint probability of coincidence is equal to $1/4$.
\end{itemize}

Could this behavior be classically simulated? Let us introduce the scenario and the strategies considered to solve this question.

\subsection{\label{sec2:subsec2} Classical setup: Uncommunicated students}
\begin{figure}[ht]
\includegraphics[width=1.\linewidth]{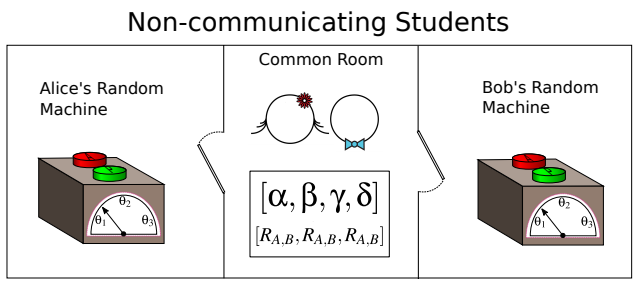}
\caption{Non-communicating students scheme. 
Two students initially find themselves in a common room where they will decide on a strategy that they will use when they become non-communicating in two distant rooms. 
In each room there is a machine that generates three random questions, which students respond by pressing two buttons, A and P. 
The common strategy chosen initially determines the potential answers to the questions of these random machines.}
\label{fig2}
\end{figure}

The experimental context for the classical situation is known in the literature as the problem of uncommunicated students~\cite{Brunner2014}. 
Basically, this \textit{gedanken} experiment is summarized in the job that has been assigned to two students who are uncommunicated (spatially separated), to reproduce the behavior of the pair of photons described before. 
Consider then two students, one of them will play the role of the photons that reach Alice, and the other will play the role of the photons that reach Bob. 
The rules are simple: both begin in the same room (the common source), after deciding how to reach their objective (common strategy) they must exit through different doors being henceforth uncommunicated, see Fig.~\ref{fig2}. Once outside the room, a question will be asked to each of them.

\begin{figure}[ht]
\includegraphics[width=0.7\linewidth]{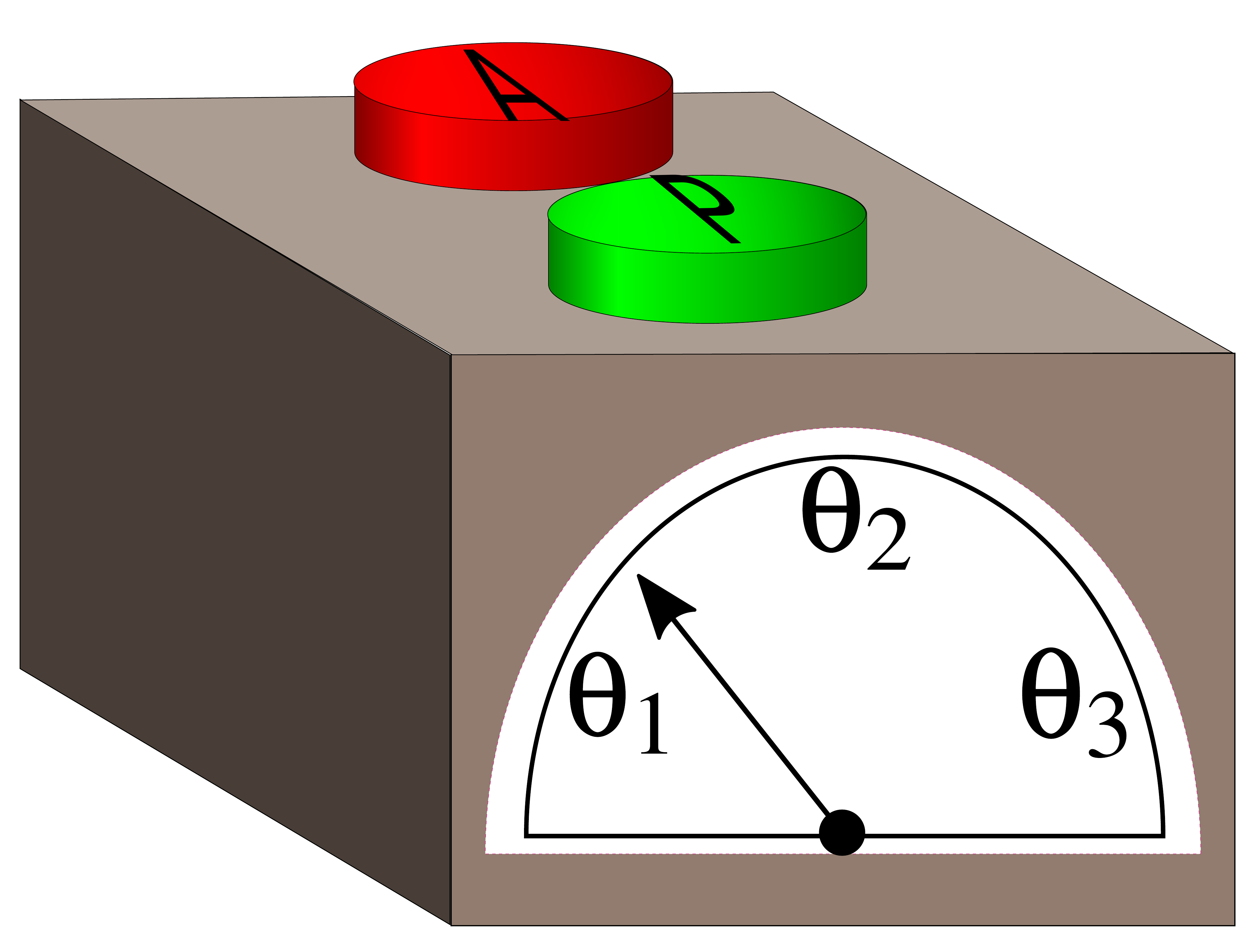}
\caption{Mermin device. When each student leaves the common room they find ahead of them this kind of device. 
In each run of experiment the machine will randomly set a question between the three options ``$0^{\circ}?$'', ``$30^{\circ}$?'', and ``$60^{\circ}$?''. Then each student should decide whether to press the $(A)$ or the $(P)$ button. The machine stores the information of the question and the answer given by the student.}
\label{fig3}
\end{figure}

What is the question? 
To answer this, suppose we have a certain device, a random machine (RM), similar to the described by N. D. Mermin~\cite{Mermin1981}. 
This device is, in essence, a random number generator which, run after run of the experiment will randomly indicate one of the possible angles, in our present case, one of three possible positions, labeled ``$0^{\circ}?$'', ``$30^{\circ}$?'', and ``$60^{\circ}$?'',  cf. Fig~\ref{fig3}. 
In addition, the device has two buttons labeled $(A)$ and $(P)$.
Each run, the students must choose either (A) or (P) and press it. 
Each device stores the information of the angle that was asked and the answer choose by the student accordingly to the following notation: $\{\theta_A \theta_B R_A R_B\}$; $A$ for Alice and $B$ for Bob. 
\Eg $\{0^{\circ} 30^{\circ} A P\}$ indicates that the device for Alice asked the question ``$0^{\circ}?$'' and the student answered $(A)$, while that device for Bob asked ``$30^{\circ}?$'' and the student answered $(P)$.

Alice and Bob have no idea of what question they will be asked every run of the experiment and once they leave the initial room they cannot communicate neither their inquiry nor her/his answer, to her/his partner. 
Nevertheless, while they are in the common room just before the run, they are allowed to build any common strategy in terms of coordinating their responses. 
Their aim is to guarantee that after many repetitions of the experiment their answers show exactly the same kind of correlations that the pair of photons that reach Alice and Bob in the quantum experiment. 
Furthermore, they are allowed to adopt a new strategy every time the experiment is repeated, so that, in the long run when the questions asked differs in $\Delta \theta$ the answers will match $\cos^2(\Delta \theta)$ of the times. 
In other words, the students should reproduce the quantum facts, \ie when they are asked for the same angle $(\Delta \theta = 0^{\circ})$, they always must give the same answer; when the questions differ by $30^{\circ}$, their answers need to agree $3/4$ of the time; and when the answers differ by $60^{\circ}$, their answers must match $1/4$ of the time~\cite{Maudlin2011quantum}.

\textit{Strategies for coincidences---} At this point, we can make two important questions: which is the best strategy for trying to reproduce the behavior of the photons? and how to know which option should they choose on each possible situation, even before both leaves the room?
In order to respond these questions, let us study all the possible combinations of the students answers---considering that they choose the same answer when asked the same question, to fulfill $F_1=1$--- by grouping such answers on 4 couples of equivalent strategies about each question. 

We will represent strategies with the following notation: $[R_{A,B}(0^{\circ}), R_{A,B}(30^{\circ}), R_{A,B}(60^{\circ})]$, where the first item represent the answer, of both students, to question ``$0^{\circ}?$'', the second to question ``$30^{\circ}?$'', and the third to question ``$60^{\circ}?$''. Therefore, the eight possible strategies are: \begin{center}
\begin{tabular}{ c c c }
	$(1) \; [P,P,P]$ & $\qquad (2) \; [A,A,A]$ & $\qquad \rightarrow \;\; \alpha$ \\
  	$(3) \; [A,P,P]$ & $\qquad (4) \; [P,A,A]$ & $\qquad \rightarrow \;\; \beta$ \\
  	$(5) \; [P,A,P]$ & $\qquad (6) \; [A,P,A]$ & $\qquad \rightarrow \;\; \gamma$ \\
    $(7) \; [P,P,A]$ & $\qquad (8) \; [A,A,P]$ & $\qquad \rightarrow \;\; \delta$ 
\end{tabular}
\end{center}

Notice that these strategies can be organized in four pairs of equivalent strategies. 
This is because we are interested in the coincidences, \ie when the questions are posed to the students, whether their results agree or not. 
It is easy to recognize that does not matters if the students chose the strategy $(1)$ or $(2)$, they always will agree even when different questions are made to them. 
In addition, if they choose the strategy $(3)$ or $(4)$ they will disagree if only the question ''$0^{\circ}$?'' is made to one of them, and they will agree in any other case, and so on.
This procedure ensures the perfect and strict correlation of fact 1, because when the same question is performed the students always choose the same answer.
Thereby, we can lump together the number of times that strategies $(1)$ and $(2)$ are played and called it the ``probability of playing $\alpha$.'' 
The proportion of times that $(3)$ and $(4)$ are played will be grouped in the ``probability of playing $\beta$,'' $(5)$ and $(6)$ will be  the ``probability of playing $\gamma$,'' and $(7)$ and $(8)$ the ``probability of playing $\delta$.''  
Therefore,  $\alpha$, $\beta$, $\gamma$, and $\delta$ must be non negative numbers, and of course this probabilities of play must satisfy: $\alpha + \beta + \gamma + \delta = 1$. 

We must remember that the students will make their decision of which strategy to adopt without information about which questions will be asked, \ie each machine generates random questions to both of them, and such process is independent of the election of the strategy that students had made. 
So in the long run, the results of this experiment will depend only on the values of $\alpha$, $\beta$, $\gamma$, and $\delta$. 
\Eg if we want to know how often the pair of questions ``$0^{\circ}?$, and ``$30^{\circ}?$'' will receive answers that disagree, then we must look for how many times the probabilities $\beta + \gamma$ were played. 

Reasoning similarly, playing the probabilities $\gamma + \delta$ give us the proportion of the experiments in which questions ``$30^{\circ}?$'', and ``$60^{\circ}?$'' will disagree, and playing the probabilities $\beta + \delta$ give us the proportion of the experiments in which questions ``$0^{\circ}?$'', and ``$60^{\circ}?$'' will disagree. 
Consequently, in order to reproduce the behavior of the photons the students must consider the probabilities such that
\begin{align*}
\beta + \gamma = 1/4 \; &\rightarrow \; \text{Disagreement} \; 0^{\circ} - 30^{\circ} \\
\gamma + \delta = 1/4 \; &\rightarrow \; \text{Disagreement} \; 30^{\circ} - 60^{\circ} \\
\beta + \delta = 3/4 \; &\rightarrow \; \text{Disagreement} \; 0^{\circ} - 60^{\circ} \end{align*}

These simple statements apparently solve the problem of the students of reproducing the behavior of the pairs of entangled photons but here is when problems become evident. 
Since, by summing $\beta+\gamma$ and $\gamma+\delta$ we obtain
\begin{equation*}
\begin{split}
(\beta + \gamma) + (\gamma + \delta) &= 1/4 + 1/4 = 1/2, \\
3/4 + 2\gamma &= 1/2,
\end{split}
\end{equation*}
where we have taken into account $\beta+\gamma$. 
It results in $\gamma = -1/8$.

However, by definition $\gamma$ must be a non negative number. 
In consequence, the conclusion is that there are not possibles scenarios in which the students could choose the strategies $(5)$ and $(6)$ $-12.5$ percent of the time. 
All this means that it does not exist a possible long-term selection of the set of strategies that students can adopt to ensure that their answers will reproduce the same correlations that photons have when prepared originally in the maximally entangled state $\ket{\phi^+}$.

\section{\label{sec3} Classical strategies for quantum coincidences}
Our goal is to introduce a strategy to determine if there is a way to simulate classically a set of quantum probabilities related to given outputs of a realistic experiment. 
Therefore, our approach it is not limited to a given initial condition although we discuss the particular case where the entangled parts are initially prepared in the quantum state $\ket{\phi^+}$. 
In order to stress the applicability of the simulation we would like to review the results considering a set of states for the photon pairs: the four Bell’s states and the two mixtures,
\begin{equation}
\begin{array}{r@{}l}
    \hat{\rho}_{Max} = &{}\frac{1}{4} \Big(\ket{HH}\bra{HH}+\ket{HV}\bra{HV} \\
     &{}+\ket{VH}\bra{VH}+\ket{VV}\bra{VV}\Big),
     \label{ecua3.1}
\end{array}
\end{equation}
and
\begin{equation}
\hat{\rho} = \frac{1}{2} \Big(\ket{HH}\bra{HH}+\ket{VV}\bra{VV}\Big).
    \label{ecua3.2}
\end{equation}

We want to discuss the possibility for the students to claim (wrongly) a successful simulation of quantum states of different nature---in the sense of the corresponding quantum facts, c.f. Table~\ref{tab:table2}.
\begin{table}[h]
\begin{ruledtabular}
\begin{tabular}{cccc}
\centering
State & \quad \; $F_1$ \quad \; & \quad \; $F_2$ \quad \; & \quad \; $F_3$ \quad \; \\
\hline
$\ket{\phi^+}$ & $1$ & $3/4$ & $1/4$ \\
$\ket{\phi^-}$ & $1/2$ & $3/8$ & $1/4$ \\
$\ket{\psi^+}$ & $1/2$ & $5/8$ & $3/4$ \\
$\ket{\psi^-}$ & $0$ & $1$ & $3/4$ \\
$\hat{\rho}_{Max}$ & $1/2$ & $1/2$ & $1/2$ \\
$\hat{\rho}$ & $3/4$ & $9/16$ & $1/4$  \\
\end{tabular}
\end{ruledtabular}
\caption{\label{tab:table2} Values of $\{F_1, F_2, F_3\}$ for all the states.}
\end{table}

In order to calculate the probabilities that in each case represent the facts we need to know the joint probabilities of coincidence of each state, c.f. Table~\ref{tab:table1} (in Appendix~\ref{AppendixA} we present the necessary steps to obtain the results of Table~\ref{tab:table1}). 
\begin{table}[ht]
\begin{ruledtabular}
\begin{tabular}{cc}
State & Probability of coincidences\\
\hline
$\ket{\phi^\pm}$ & $\cos^2(\theta_s \mp \theta_i)$ \\
$\ket{\psi^\pm}$ & $\sin^2(\theta_s \pm \theta_i)$ \\
$\hat{\rho}_{Max}$ & $\frac{1}{2}$ \\
$\hat{\rho}$ & $\frac{1}{2}\left[\cos(2\theta_s)\cos(2\theta_i) + 1\right]$
\end{tabular}
\end{ruledtabular}
\caption{\label{tab:table1} Probabilities of Coincidence for the four Bell's states and the two statistical mixtures $\hat{\rho}_{Max}$ and $\hat{\rho}$.}
\end{table}

Once we know the expressions for the probability of coincidence, we can calculate the value of the probabilities for each of the quantum facts.
In Table~\ref{tab:table2}, we present a recap of the values of $\{F_1, F_2, F_3\}$ for all the states.
The previous analysis of probabilities that lead to a negative $\gamma$ coefficient, in sec. II B, was based on the mismatches, \ie $F_2$ and $F_3$. 
It was obtained also imposing the strict condition of fulfilling always $F_1=1$ which is set by our initial condition, $\ket{\phi^+}$.
When this condition it is established and the analysis is done taking into account just the disagreements, what we obtain is $\gamma = -1/8,1/2,1/4,-1/8,1/4,1/16$ for the states $\ket{\phi^+}, \ket{\phi^-}, \ket{\psi^+} \ket{\psi^-}, \hat{\rho}_{Max}$, and $\hat{\rho}$ respectively.
Notice that the only states that have a negative $\gamma$ are $\ket{\phi^+}$ and $\ket{\psi^-}$.
These have $F_1$ equals 1 and 0, respectively.
From the point of view of the correlations they are equivalent because having total certainly or total ignorance relates exactly to the same correlation.
Therefore, they are of the same nature in the sense of quantum facts.

We show with our simulation that in the subspace of disagreements, $F_2$ vs $F_3$, is impossible for the students to obtain the probabilities that characterize the states $\ket{\phi^+}$ and $\ket{\psi^-}$ but in the case of $\ket{\phi^-}, \ket{\psi^+}, \hat{\rho}_{Max}$, and $\hat{\rho}$ they could reproduce their facts and erroneously conclude that they indeed simulate a quantum state.

\subsection{\label{sec3:subsec1} Description of simulation}
We propose a numerical simulation \cite{renosimul} that runs through the entire parameter space $\alpha$, $\beta$, $\gamma$ and $\delta$, by discrete variations of a value $\Delta p$ of each parameter. 
We guarantee that the condition $\alpha + \beta + \gamma + \delta = 1$  it is always satisfied, that is,  certifying that we cover all the possible strategies that students can choose. 
\Eg if $\Delta p = 0.1$ and we start at configuration $[1.0,0.0,0.0,0.0]$, the next set of strategies will be $[0.9,0.1,0.0,0.0]$, and the next $[0.9,0.0,0.1,0.0]$, and so on until the last term to be evaluated is $[0.0,0.0,0.0,1.0]$. 
Varying in this way the parameters, they will form a set which has a size defined by the tetrahedral number
\begin{equation}
T_n = \displaystyle\binom{n + 2}{3},
\label{ecua3.3}
\end{equation}
where $n = (1/\Delta p) + 1$, and $\Delta p = 1/p$ with $p \in \mathbb{N}$. 
Consider again the case of $\Delta p = 0.1$, here we have $n = 11$ and $T_n = 286$, which is the size of the grid of the entire set of parameters. 
Evidently, as smaller is $\Delta p$, bigger is $T_n$, and bigger is the size of the set of probabilities that define the amount of different experiments that the simulation will perform. 
So, this number is the first important factor in the simulation conditioned by $\Delta p$. 

Furthermore, for each selection of $\alpha$, $\beta$, $\gamma$ and $\delta$, the simulation will run the experiment a great number of times in order to verify if that selection satisfy fact 2 and fact 3 simultaneously. 
Then, $\alpha$ times of the experiments the simulation will choose randomly between strategy $(1)$ and strategy $(2)$, $\beta$ times the simulation will choose randomly between strategy $(3)$ and strategy $(4)$, $\delta$ between strategy $(5)$ and strategy $(6)$, and $\delta$ between strategy $(7)$ and strategy $(8)$. 
\Eg the selection $[1,0,0,0]$ means that the simulation choose randomly between strategies $(1)$ and $(2)$ $100$ percent of the time, or equivalently, $50\%$ of the time simulation will choose strategy $(1)$, and the other $50\%$ of the time strategy $(2)$. 
Note that we cover the whole space of parameters with a distance $\Delta p$ in all directions between point and point. Therefore, reaching an accuracy of $\Delta p$.

\subsection{\label{sec3:subsec2} Uncommunicated students results}
We present the results of simulation for $F_2$ and $F_3$ together with the probabilities of coincidence for all the Bell’s states, $\ket{\phi^\pm}$ and $\ket{\psi^\pm}$, and the mixtures, $\hat{\rho}_{Max}$ and $\hat{\rho}$ considering the space of parameters $(\alpha, \beta, \gamma)$ for a fixed $\Delta p$. 
Note that we only need three of the parameters since $\delta = 1 - (\alpha + \beta + \gamma)$.

\begin{figure}[ht]
\includegraphics[width=1.\linewidth]{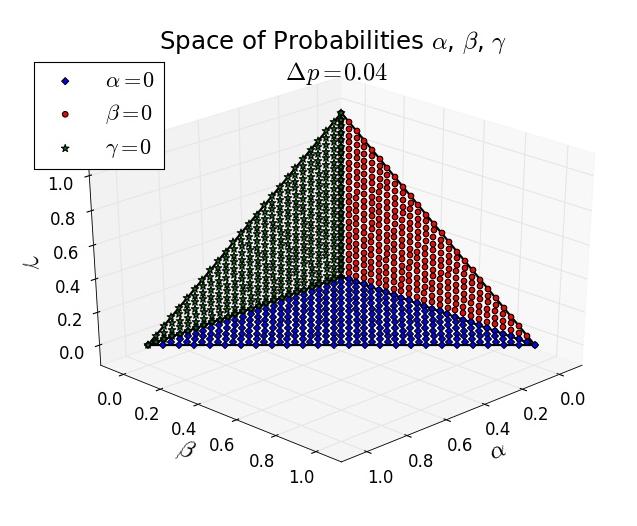} 
\caption{Parameters space $(\alpha, \beta, \gamma)$, restricted to the planes $\alpha = 0$ (blue), $\beta = 0$ (red), and $\gamma = 0$ (green). Each point is mapped into the facts space $(F_2, F_3)$.}
\label{fig_4}
\end{figure}

\begin{figure}[ht]
	\includegraphics[width=1.0\linewidth]{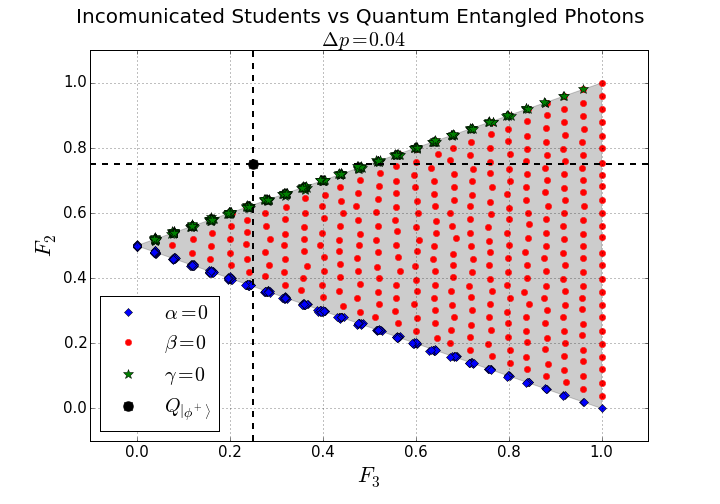}
	\caption{Facts space corresponding to each point in parameters space. Note that the Q point represents the quantum prediction for the probability of coincidence of the state $\ket{\phi^+}$.}
	\label{fig_5}
\end{figure}

The parameters space is given by a regular tetrahedron, this is a geometrical imposition over the space of classical probabilities. 
To each point of the space $(\alpha, \beta, \gamma)$ corresponds a set of facts $(F_1,F_2,F_3)$. 
On account of the design of the protocol the fact 1 is always satisfied, which implies that $F_1 = 1$ independently of the set of chosen values $(\alpha, \beta, \gamma)$. 
Therefore, each point of space $(\alpha, \beta, \gamma)$ is mapped into a $(F_2,F_3)$ plane. 
In Fig.~\ref{fig_4}, we show the planes in the parameters space corresponding to configuration $\alpha = 0$, $\beta = 0$, and $\gamma = 0$. 
Note that, the planes $\alpha = 0$ and $\gamma = 0$ are mapped into the triangular frontier of classical facts space, c.f. Fig.~\ref{fig_5}. 
Furthermore, we show the point ``Q'' which corresponds to the quantum prediction for the probability of coincidence of the entangled state $\ket{\phi^+}$. 

\begin{figure}[ht]
\includegraphics[width=1.\linewidth]{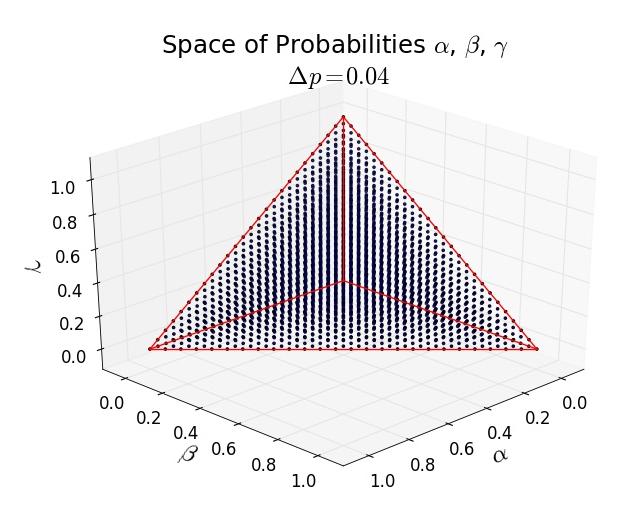}
\caption{Parameters space over the whole parameters values with distance $\Delta p = 0.04$ between the points.}
\label{fig_6}
\end{figure}

\begin{figure}[ht]
\includegraphics[width=1.0\linewidth]{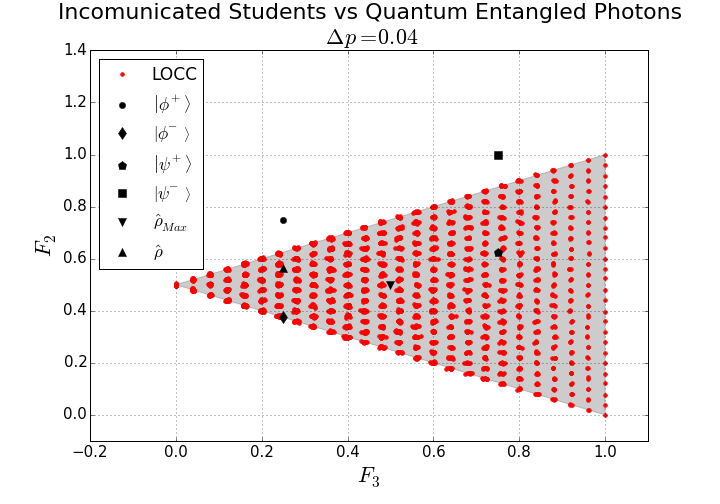}
\caption{Parameters space mapped into the facts space, besides, we show the points for the other quantum states: $\ket{\phi^-}$, $\ket{\psi^{\pm}}$, $\hat{\rho}_{Max}$, and $\hat{\rho}$.}
\label{fig_7}
\end{figure}

In Fig.~\ref{fig_6} we show the result of a sweep over the whole space of classic parameters with a fixed distance between each point, that is, $\Delta p = 0.04$. 
Again, for each point in space $(\alpha, \beta, \gamma)$ we obtain one point in facts space $(F_2,F_3)$, see Fig.~\ref{fig_7}. 
It is important to note that the classical space have a geometrical restriction that is violated by the probability of coincidence of state $\ket{\phi^+}$. 
Such geometrical restriction can be established from the two straight lines that confine the classical region, that is,
\begin{subequations}
\label{ecua3.4}
\begin{equation}
F_2 = \frac{1}{2}(F_3 + 1),
\label{ecua3.4.a}
\end{equation}
\begin{equation}
F_2 = -\frac{1}{2} (F_3 - 1).
\label{ecua3.4.b}
\end{equation}
\end{subequations}

These last equations define the maximal and minimal value of $F_2$ in the classical region, respectively. 
Mathematically it is, $-\frac{1}{2}(F_3 - 1) \leq F_2 \leq \frac{1}{2}(F_3 + 1)$ or equivalently $-F_3 \leq 2 F_2 - 1 \leq F_3$.
From where we can obtain the inequality imposed by the classical restrictions, given by
\begin{equation}
|2 F_2 - 1| \leq F_3.
\label{ecua3.5}
\end{equation}

According to our numerical simulation, any strategy, or any choice, that uncommunicated students can make it is included in such classical region. 
Clearly, the point "Q", corresponding to $\ket{\phi^+}$ is outside of such region, showing thus, that there is no classical strategy capable of reproducing the behavior the pairs of entangled photons.

In Fig.~\ref{fig_7} we also show the quantum predictions for the other states. 
The red points represent all the states the students can simulate with their classical strategies within the whole volume of the tetrahedron.
All possible results of a classical protocol.
The black points describe the given quantum states.
The states $\ket{\phi^-}$, $\ket{\psi^+}$, $\hat{\rho}_{Max}$, and $\hat{\rho}$ appear to be contained in the classical region, but, remember that such states do not satisfy the condition $F_1 = 1$ imposed by the initial state $\ket{\phi^+}$. 
As expected $\ket{\phi^+}$ is not the only state that has its point out of the classical region.
The equivalency between $F_1=1$ and $F_1=0$, having total certainly or total ignorance, makes $\ket{\psi^-}$ also impossible to simulate with the set of strategies used by the students that could reproduce its behavior in the ($F_2-F_3$) plane.
The results show that the students may believe they simulated quantum states, even entangled, but that is a mere consequence of not considering all the quantum facts.
The protocol can be modified in order to satisfy each value of $F_1$ for each quantum state.
States with equivalent $F_1$ will be out of the classical zone and the rest could be wrongly reported to be successfully simulated via classical strategies due to the lack of information.

\section{\label{sec4} Discussion and conclusions}
Quantum effects cannot be classically simulated.
That said, we conclude that an entangled state have some partial correlations that can be always reproduced classically. 
In this case, correlations associated to fact $2$ and fact $3$ can be reproduced, but for the proper characterization of the entangled state more correlations are needed, given by fact $1$.
Therefore, when all the geometrical conditions imposed by quantum correlations are taken into account is not possible to reproduce solely with classical strategies the corresponding probabilities that characterize any quantum state.
Additionally, for any entangled state it is possible to find other different directions of polarization that produce facts that can not be emulated by non-communicating students. 
In this sense, the key property is the entanglement and the geometric constraint imposed by Bell's theorem on classical results.

We have carried out an exhaustive search of all the possible classical strategies that try to reproduce the correlations of a pair of entangled photons, and we have found that non of these strategies can reproduce the quantum predictions. The procedures performed by non-communicating students are a proper implementation of LOCC, \ie students can only respond to each experiment in their own local environment, outside the common room which is the only space where they can communicate and choose an strategy. 

The conclusion is that no classical procedure LOCC is capable to reproduce the quantum entangled states. 

On the other hand, the geometric constraint that we have found on the classical facts space must naturally be related to some kind of Bell inequality, a relation that we are still studying.

\appendix
\section{Quantum Description of Probability of Coincidences}
\label{AppendixA}
In order to describe the probability of coincidence for a pair of entangled photons in polarization we describe the physical system in the polarization basis, \{$H$, $V$\}. 
Here we calculate the joint probability of detecting a signal photon and idler photon to have elliptical polarization states.

\subsection{Probabilities and Operators}
\label{AppendixA:subsec1} 
First, we are going to define the general state of the \{$H$, $V$\} basis which have the form:
\begin{equation}
\ket{\psi} = c_1 \ket{H} + c_2 \ket{V},
\label{ecuaA.1}
\end{equation}
where $c_1$, $c_2$ $\in \mathbb{C} $, must satisfy $|c_1|^2 + |c_2|^2 = 1$. 
This means we can write $c_1$ and $c_2$ like this:
\begin{subequations}
\label{ecuaA.2}
\begin{equation}
	c_1 = |c_1| \; e^{i \theta_1},
    \label{ecuaA.2.a}
\end{equation}
\begin{equation}
	c_2 = |c_2| \; e^{i \theta_2},
    \label{ecuaA.2.b}
\end{equation}
\end{subequations}
with $|c_1| = \cos(\theta)$, and $|c_2| = \sin(\theta)$. 
Replacing these in Eq.~(\ref{ecuaA.1}) we obtain 
\begin{equation}
\begin{split}
	\ket{\psi} &= \cos(\theta)\;e^{i \theta_1}\ket{H} + \sin(\theta)\;e^{i \theta_2}\ket{V} \\
			   &= \cos(\theta)\;\ket{H} + \sin(\theta)\;e^{i\varphi}\ket{V},
	\label{ecuaA.3}
\end{split}
\end{equation}
where $\varphi = \theta_2 - \theta_1$, and the global phase $e^{i \theta_2}$ is deprecated. 
Hence, we can write the general states for elliptical polarization and their respective orthogonal states as:
\begin{subequations}
\label{ecuaA.4}
\begin{equation}
	\ket{e_{s,i}} = \cos(\theta_{s,i})\ket{H} + \sin(\theta_{s,i})\;e^{i\varphi_{s,i}}\ket{V}, 
	\label{ecuaA.4.a}
\end{equation}
\begin{equation}
	\ket{e_{s,i}^{\perp}} = \sin(\theta_{s,i})\ket{H} - \cos(\theta_{s,i})\;e^{-i\varphi_{s,i}}\ket{V}.
    \label{ecuaA.4.b}
\end{equation}
\end{subequations}

Here, the sub-indexes $s$ and $i$ means signal photon and idler photon, respectively, and the super index $\perp$ means orthogonal. 
With the first two states in the Eq.~(\ref{ecuaA.4.a}) we can write the projection operator. 
And equivalently, with the last two states in Eq.~(\ref{ecuaA.4.b}) we can write the projection operator for the orthogonal states as
\begin{subequations}
\label{ecuaA.5}
\begin{equation}
	\hat{P}(e_s,e_i) = \ket{e_s,e_i}\bra{e_s,e_i},
    \label{ecuaA.5.a}
\end{equation} 
\begin{equation}
	\hat{P}(e_s^{\perp},e_i^{\perp}) = \ket{e_s^{\perp},e_i^{\perp}}\bra{e_s^{\perp},e_i^{\perp}}.
	\label{ecuaA.5.b}
\end{equation} 
\end{subequations}

Note that, we can write these two operators in terms of the \{$H$, $V$\} basis, but such work is excessively long and inefficient resulting in sixteen terms. 
These operators in Eqs.~(\ref{ecuaA.5}) allow us to calculate the joint probability that a given state has elliptical polarization, or orthogonal to it, as the case may be. 
Such joint probability is defined as
\begin{subequations}
\label{ecuaA.6}
\begin{equation}
P_{\ket{\psi}}(e_s,e_i) = |\bra{\psi}\ket{e_s,e_i}|^2,
\label{ecuaA.6.a}
\end{equation}
\begin{equation}
P_{\ket{\psi}}(e_s^{\perp},e_i^{\perp}) = |\bra{\psi}\ket{e_s^{\perp},e_i^{\perp}}|^2.
\label{ecuaA.6.b}
\end{equation}
\end{subequations}

However, these joint probabilities, Eq.~(\ref{ecuaA.6}), will not give us the complete set of coincidences in an experiment as described in Section \ref{sec2:subsec1}. 
To obtain the probability of coincidences, it is necessary to take into account the probability that both photons are detected, \ie they both pass through the polarizer; as well as the probability that they will not pass, \ie they are not detected. 
Such probability can be written as the sum of the probability of measurement (or detection) the angles $\theta_s$ and $\theta_i$ plus the probability of measurement the orthogonal angles $\theta_s^{\perp}$ and $\theta_i^{\perp}$, that is
\begin{equation}
\begin{split}
P_{\ket{\psi}}^{C}(\theta_s,\theta_i) &= P_{\ket{\psi}}^{D}(\theta_s,\theta_i) + P_{\ket{\psi}}^{\overline{D}}(\theta_s,\theta_i) \\
&= P_{\ket{\psi}}^{D}(\theta_s,\theta_i) + P_{\ket{\psi}}^{D}(\theta_s^{\perp},\theta_i^{\perp}),
\label{ecuaA.7}
\end{split}
\end{equation}
where $C$, $D$ and $\overline{D}$ means \textit{coincidences}, \textit{detection}, and \textit{not detection}, respectively. 
Note that the last term in Eq.~(\ref{ecuaA.7}) show to us that the probability of no detection of the state in angles $\theta_s$ and $\theta_i$ is the same that the probability of detection of the state in the orthogonal angles $\theta_s^{\perp}$ and $\theta_i^{\perp}$.

In the rest of this section we are going to calculate such joint probability of coincidences for two kind of states: the first group of states are the four Bell's entangled states~\cite{Beck2012quantum}, the second one are two particular mixed states; which are described as
\begin{subequations}
	\label{ecuaA.8}
\begin{equation}
	\ket{\phi^\pm} = \frac{1}{\sqrt{2}}\left(\ket{HH} \pm \ket{VV}\right),
	\label{ecuaA.8.a}
\end{equation}
\begin{equation}
	\ket{\psi^\pm} = \frac{1}{\sqrt{2}}\left(\ket{HV} \pm \ket{VH}\right),
    \label{ecuaA.8.b}
\end{equation}
\end{subequations}
and
\begin{widetext}
\begin{subequations}
	\label{ecuaA.9}
\begin{equation}
	\hat{\rho}_{Max} = \frac{1}{4} \left[\ket{HH}\bra{HH}+\ket{HV}\bra{HV}+\ket{VH}\bra{VH}+\ket{VV}\bra{VV}\right],
	\label{ecuaA.9.a}
\end{equation}
\begin{equation}
	\hat{\rho} = \frac{1}{2} \left[\ket{HH}\bra{HH}+\ket{VV}\bra{VV}\right],
    \label{ecuaA.9.b}
\end{equation}
\end{subequations}
\end{widetext}
respectively. 

\subsubsection{\label{sec3:subsec1:subsubsec1} For Bell's states}
First, we are going to calculate the joint measurement probabilities $P(e_s,e_i)$ and $P(e_s^{\perp},e_i^{\perp})$, Eq.~(\ref{ecuaA.6}), for photons prepared in each one of Bell's states, Eq.~(\ref{ecuaA.8}), starting from $\ket{\phi^+}$, that is
\begin{align}
	P_{\ket{\phi^+}}(e_s,e_i) &= \bra{\phi^+}\ket{e_s,e_i}\bra{e_s,e_i}\ket{\phi^+} \\
    = \frac{1}{2}& |\bra{HH}\ket{e_s,e_i}+\bra{VV}\ket{e_s,e_i}|^2 \\
    = \frac{1}{2}& |\cos(\theta_s)\cos(\theta_i) \\
    &+ e^{i(\varphi_s + \varphi_i)}\;\sin(\theta_s)\sin(\theta_i)|^2.
\label{ecuaA.10}
\end{align}

The result of Eq.~(\ref{ecuaA.10}) is for a general elliptical polarization joint probability of pairs of single photons prepared in the state $\ket{\phi^+}$. But we are interested only in linear polarization, \ie $\varphi_s = \varphi_i = 0$, that means that the probability only depends of angles $\theta_s$ and $\theta_i$, so we obtain
\begin{align}
P_{\ket{\phi^+}}(\theta_s,\theta_i) &= \frac{1}{2}\; |\cos(\theta_s)\cos(\theta_i) \\ 
                    &+ \sin(\theta_s)\sin(\theta_i)|^2 \\
					&= \frac{1}{2}\;|\cos(\theta_s-\theta_i)|^2 \\
                    &= \frac{1}{2}\;\cos^2(\theta_s-\theta_i).
\label{ecuaA.11}
\end{align}

Similarly, for the orthogonal projection operator in Eq.~(\ref{ecuaA.6}) we obtain the joint probability
\begin{equation}
P_{\ket{\phi^+}}(\theta_s^{\perp},\theta_i^{\perp}) = \frac{1}{2}\;\cos^2(\theta_s-\theta_i).
\label{ecuaA.12}
\end{equation}

This means that the probability of coincidence, Eq.~(\ref{ecuaA.7}), \ie the probability of measurement (or detection) the photons in the angles $\theta_s$ and $\theta_i$ plus the probability of measurement the photons in the orthogonal angles $\theta_s^{\perp}$ and $\theta_i^{\perp}$ for the prepared state $\ket{\phi^+}$, have the form
\begin{equation}
\begin{split}
P_{\ket{\phi^+}}^{C}(\theta_s,\theta_i) &= \frac{1}{2}\;\cos^2(\theta_s-\theta_i) \\
&+ \frac{1}{2}\;\cos^2(\theta_s-\theta_i) \\
&= \cos^2(\theta_s-\theta_i).
\label{ecuaA.13}
\end{split}
\end{equation}

The Eq.~(\ref{ecuaA.13}) show us the perfect correlation of state $\ket{\phi^+}$, \ie when the angles of polarizations of Alice and Bob are the same, then the probability of coincidence is always equal to one. For the other three Bell's states: $\ket{\phi^-}$, $\ket{\psi^+}$ and $\ket{\psi^-}$,  we obtain the following probabilities of coincidence.

The Bell's state $\ket{\phi^-}$ has probability of coincidence
\begin{equation}
P_{\ket{\phi^-}}^{C}(\theta_s,\theta_i) = \cos^2(\theta_s + \theta_i).
\label{ecuaA.14}
\end{equation}

The Bell's state $\ket{\psi^+} $ has probability of coincidence
\begin{equation}
P_{\ket{\psi^+}}^{C}(\theta_s,\theta_i) = \sin^2(\theta_s + \theta_i).
\label{ecuaA.15}
\end{equation}

The last but not least of Bell's states $\ket{\psi^-}$ has probability of coincidence
\begin{equation}
P_{\ket{\psi^-}}^{C}(\theta_s,\theta_i) = \sin^2(\theta_s - \theta_i).
\label{ecuaA.16}
\end{equation}

\subsubsection{\label{sec3:subsec1:subsubsec2} For mixed states}
Additional to Bell's states, we are interested in the two particular mixed states in Eq.~(\ref{ecuaA.9}). Note that the first one is a maximally mixed state, and the other one is a partially mixed state.

The probabilities of coincidence for such states are
\begin{equation}
\begin{split}
P_{\hat{\rho}_{Max}}^{C}(\theta_s,\theta_i) = \frac{1}{2},
\label{ecuaA.17}
\end{split}
\end{equation}
and
\begin{equation}
\begin{split}
P_{\hat{\rho}}^{C}(\theta_s,\theta_i) = \frac{1}{2} \left[\cos(2\theta_s)\cos(2\theta_i) + 1\right],
\label{ecuaA.18}
\end{split}
\end{equation}
respectively. Note that $P_{\hat{\rho}_{Max}}^{C}(\theta_s,\theta_i)$ in Eq.~(\ref{ecuaA.17}) is independent of the angles $\theta_s$ and $\theta_i$, and is always equal to $1/2$. A resume of all the joint probabilities of coincidence calculated are show in table~\ref{tab:table1}.

\end{document}